\documentclass[aps,pre,preprint,superscriptaddress,showpacs]{revtex4}

\usepackage{graphicx}
\usepackage{amsfonts}
\newcommand{\myref}[1]{(\ref{#1})}

\newcommand{\kb}{k} 

\newcommand{\mucoex}{\mu_{\mathrm{coex}}} 

\newcommand{\Peclet}{\mathrm{Pe}} 
\newcommand{\Weber}{\mathrm{We}} 
\newcommand{\Reynolds}{\mathrm{Re}} 
\newcommand{\Cahn}{\mathrm{C}} 
\newcommand{\Capillary}{\mathrm{Ca}} 

\newcommand{\unit}[1]{\tilde{#1}}
  
\newcommand{\shear}{\dot{\gamma}}  
\newcommand{\bfr}{\mathbf{r}}  
\newcommand{\diffd}{\mathrm{d}}  

\newcommand{\molf}{\varphi}
\newcommand{\stene}{\mathbf{e}}
\newcommand{\norm}[1]{\|#1\|}
\newcommand{\disnab}{\hat{\nabla}}
\newcommand{\ptilde}{\tilde{p}}
\newcommand{\CapillaryTube}{\Capillary_\mathrm{T}}
\newcommand{\CapillaryCritical}{\Capillary_\mathrm{c}}

\begin{document}

\title{Modeling of droplet breakup in a microfluidic T--shaped
junction with a phase--field model}

\author{Mario De Menech}
\affiliation{Max--Planck Institut f\"ur Physik komplexer Systeme, \\
N\"othnitzer Str. 38, 01187 Dresden, Germany}
\affiliation{Theoretische Physik, Fachbereich Naturwissenschaften, 
Universit{\"a}t Kassel\\ 
Heinrich-Plett-Str.\,40, 34132 Kassel, Germany}
\affiliation{Unilever R\&D Vlaardingen\\ 
Olivier van Noortlaan 120, 3133 AT Vlaardingen, The Netherlands}

\date{\today}

\begin{abstract}
A phase--field method is applied to the modeling of flow and breakup of
droplets in a T--shaped junction in the hydrodynamic regime where capillary
and viscous stresses dominate over inertial forces, which is characteristic of
microfluidic devices. The transport equations are solved numerically in the
three--dimensional geometry, and the dependence of the droplet breakup on the
flow rates, surface tension and viscosities of the two components is
investigated in detail. The model reproduces quite accurately the phase
diagram observed in experiments performed with immiscible fluids.  The
critical capillary number for droplet breakup depends on the viscosity
contrast, with a trend which is analogous to that observed for free isolated
droplets in hyperbolic flow.
\end{abstract}


\maketitle

\section{Introduction}

\begin{figure}[htb]
\begin{center}
\includegraphics[width=0.37\textwidth]{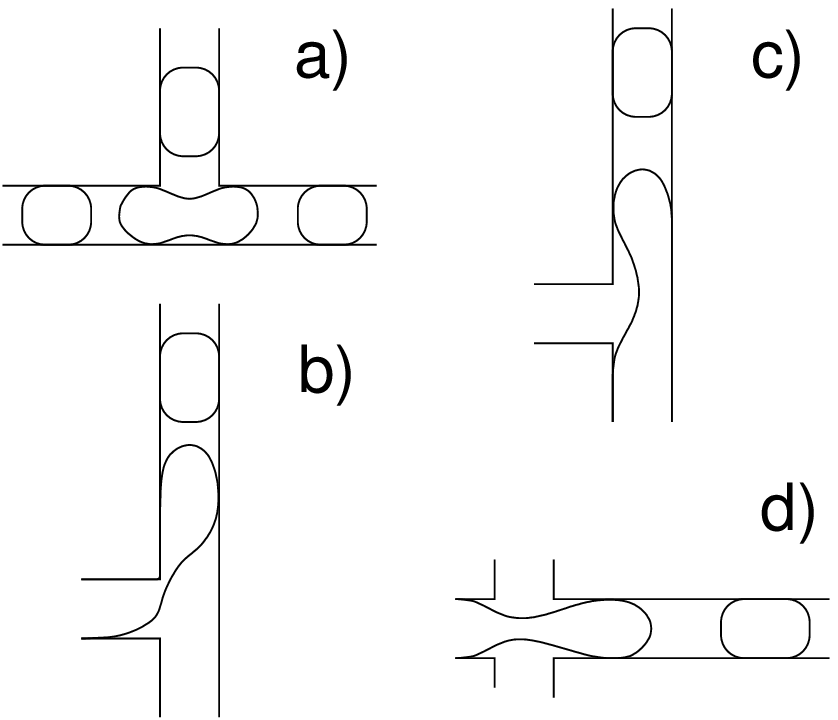}
\includegraphics[width=0.08\textwidth]{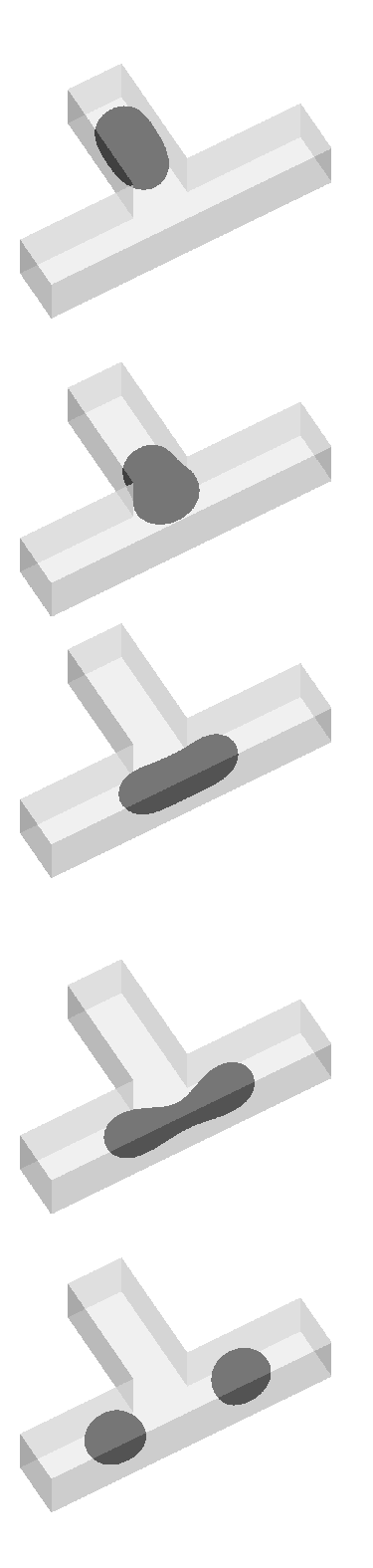}
\end{center}
\caption{Left: examples of simulations of two--phase flow dynamics in
  microfluidic devices: a) droplet breakup in a symmetric T--junction; b) and
  c), droplet formation in a T--junction with cross flow; d) droplet formation
  in a flow--focusing cross--junction. Right: frames showing the modeled
  three--dimensional flow of a droplet in a symmetric T--junction.}
\label{fig:devices}
\end{figure}

Much of the recent interest in the behavior of fluids in small channels stems
from the wide range of applications that microfluidic devices are finding in
biology and chemistry. The laminar regime of the flow in these systems allows
an accurate control of mass and momentum transport, such that chemical species
can be brought together in small quantities within restricted regions of
space, enabling, for example, fast mixing~\cite{Knight} and high
resolution of reaction kinetics~\cite{Baroud}. In the case of two--phase
systems, nano-- and pico--liter droplets can be handled with great precision
when the destabilizing effects of wall impurities can be suppressed, producing
well controlled structures~\cite{Anna, Link, Ganan-Calvo,Thorsen, Dreyfus};
partial wetting, in fact, introduces non-linear instabilities which may
generate complex erratic \cite{Dreyfus} or oscillatory
patterns~\cite{Kuksenok2003}. The variety of two--phase flow behaviors (see
figure \ref{fig:devices}) is both interesting at a fundamental level and
relevant for those applications based on the accurate tuning of droplet
sizes~\cite{Joanicot2005}: micro--droplets could, for example, be used to
confine chemical reactions, or to deliver substances in well defined
amounts. Such a control is generally attainable for a limited range of flow
conditions and with fluids having properties that are compatible with the
surfaces of the device used.  The combinations of materials, geometries and
flow conditions which are being used in experimental setups are rapidly
growing, and modeling can be extremely helpful both in providing insight and
in improving the design and prototyping of microfluidic devices. In this
respect, the main challenge of a modeling technique is to cope with the
three--dimensionality and the strong
surface stresses present both at the fluid--fluid and fluid--solid interfaces.

Phase--field models have been developed for the study of the dynamics of
critical phenomena, in particular in relation to nucleation and spinodal
decomposition~\cite{CahnSpinodal1965}. The spatial variations of the phase
order parameter indicate the location of interfaces separating the growing
domains, and interface coalescence and breakup is driven by the local
curvature--dependent solubility (Gibbs--Thomson effect): interfacial regions
of high curvature have a higher chemical potential which drives a mass flux to
the surrounding medium to reduce the curvature and therefore the chemical
potential gradients. The effects of fluid flow are included by adding an
advection term to the diffusion equation of the order parameter, which is then
coupled to a momentum transport equation to describe the hydrodynamics of a
multiphase system~\cite{HohenbergReview1977,Anderson1998}. From a computational
point of view, modeling of fluid boundaries as having finite thickness greatly
simplifies the handling of interface coalescence or breakup events, and in the
case of phenomena occurring near the critical point, this approach is fully
justified by the fact that, in the mean field approximation, interfaces are
indeed diffuse, with a characteristic length which is comparable with the
typical domain size. On the other hand, it can be shown that phase--field
physics approaches asymptotically that of two immiscible fluids as the
interface thickness becomes small compared to the local curvature, both for
unbounded flow and in the presence of solid
surfaces~\cite{Chella1996,Jacqmin-NS,Jacqmin-contact}.

In this paper we use a phase--field model to study the three--dimensional flow
 of a binary mixture in constrained geometries in the hydrodynamic regime
 which characterizes the flow in microfluidic devices, i.e. with low Reynolds
 and Capillary numbers; we show how the computed droplet flow and breakup
 dynamics in a T--shaped junction follows closely that observed experimentally
 for water in oil mixtures, despite the fact that the interface thickness in
 the simulations is far from the sharp--interface limit.
%
%

The paper is structured as follows: section \ref{sec:model} summarizes how the
 thermodynamic and transport properties of the non--homogeneous system can be
 derived from a free--energy functional; section \ref{sec:method} describes
 the numerical method used to solve the transport equations while section
 \ref{sec:free_flow} presents the free--flow tests which were performed to
 validate the implementation of the code. Finally, in sections
 \ref{sec:channels} and \ref{sec:Tjunction} the results for the modeling of
 droplet flow and breakup in a T--shaped junction are reported.
%
%

\section{Model equations}
\label{sec:model}

A unifying feature of all phase--field models is the existence of a
free--energy functional, which, besides determining the equilibrium
properties, strongly influences the dynamics of the system, whose transport
equations can be derived from conservation arguments~\cite{Anderson1998}, or
from generalized hydrodynamic theory~\cite{ChaikinBook}. In this section we
briefly review this theoretical framework for a two--phase system,
illustrating in particular how the surface tension, the wetting properties and
the pressure tensor follow consistently from the choice of the free--energy
functional.

\subsection{Thermodynamics of the non--homogeneous system}

We consider a mixture of two fluids, $A$ and $B$, with the Cahn--Hilliard--van
der Waals form for the free energy in the volume $V$, 
\begin{equation}
F[n,\molf]=\int_V \diffd\bfr\; \left\{
\frac{\kappa n}{2} |\nabla \molf|^2 + n W(\molf) + f(n) \right\},
\label{eq:free_energy}
\end{equation}
where $n=n_A+n_B$ is the total particle number density, $\molf=n_B/n$ is the
molar fraction of component $B$, $\kappa$ determines the surface tension (see
section \ref{sec::surface_tension} below) and $f$ is the sum of the
free--energy densities of the pure components, which for simplicity is assumed
to depend only on the total density.  Also, $W$ is the free energy of mixing,
for which we shall take the expression for a simple symmetric
mixture~\cite{PrausnitzBook},
 \begin{equation}
 W(\molf)=2 k T_c \molf (1-\molf) + kT \left[\molf \log 
 \molf+(1-\molf) \log(1-\molf)\right],
\label{eq:simple_mixture}
\end{equation}
which is the sum of the ideal and the excess parts, $\kb$ is Boltzmann's
constant, and $T$ and $T_c$ are, respectively, the absolute and critical
temperatures. The critical point corresponds to the molar fraction
$\molf_c=1/2$ and temperature $T_c$, so that below $T_c$ the system phase
separates; at $T=0$ the two components completely demix so that the
equilibrium bulk molar fractions become $\molf_A=0$ and $\molf_B=1$. 
In the incompressible limit, $n=$const., the equilibrium concentration
satisfies the Euler--Lagrange equation
\begin{equation}
\mu(\bfr)=-\kappa \nabla^2 \molf(\bfr)+W^\prime(\molf(\bfr))=\mucoex,
\label{eq:EulerLagrange}
\end{equation}
where $\mu(\bfr)=\delta F/\delta\molf(\bfr)$ is the chemical potential
difference, and $\mucoex$ is the Lagrange multiplier which fixes the total
molar fraction.  For a flat interface $\mucoex=0$ and $\molf$ is a function of
one spatial coordinate only: the concentration profile varies between the
equilibrium molar fractions $\molf_A$ and $\molf_B$, within a region having a
finite characteristic thickness $\xi$ proportional to $\sqrt{\kappa/k
T_c}$. In non--equilibrium conditions, local
imbalances of the chemical potential $\mu(\bfr)$ will generate diffusion
currents which tend to restore the configuration satisfying equation
\myref{eq:EulerLagrange}.

\subsection{Surface tension}
\label{sec::surface_tension}

The mechanical definition of surface tension for a flat interface is
given by the integral of the difference between the normal and
transverse pressures:
\begin{equation}
\sigma=\int_{-\infty}^{\infty} dx \;\left(P_{xx}- \frac{P_{yy}+P_{zz}}{2}
\right),
\label{eq:sigma}
\end{equation}
where $x$ is the direction normal to the interface.  The pressure
tensor must satisfy the Gibbs--Duhem equation (we consider only
isothermal transformations)
\begin{equation}
\nabla_\beta \cdot P_{\alpha \beta} = n_A\nabla_\alpha \mu_A + n_B
\nabla_\alpha \mu_B,
\label{eq:Gibbs-Duhem}
\end{equation}
where $\mu_A=\delta F/\delta n_A$ and $\mu_b=\delta F/\delta n_B$ are the
chemical potentials computed as functional derivatives of the free
energy~\myref{eq:free_energy}. The integrand in the
functional~\myref{eq:free_energy} is a Lagrangian density, and is invariant
under space translations; Noethers theorem can be applied to construct the
expression for the associated conserved current~\cite{Anderson1998}, leading
to the explicit form
\begin{equation}
P_{\alpha \beta} = \delta_{\alpha\beta} \left[ n f^\prime(n) -f(n)
\right] +\kappa n \nabla_\alpha \molf \nabla_\beta \molf,
\label{eq:pressure-tensor}
\end{equation}
where we recognize the non--isotropic part (the second term
in~\myref{eq:pressure-tensor}) first introduced by Korteweg.  From
equations~\myref{eq:sigma} and~\myref{eq:pressure-tensor}, we get
\begin{equation}
\sigma=\int_{-\infty}^{\infty} dx \;\kappa n \left(\frac{d
\molf}{dx}\right)^2,
\label{eq:surface-tension}
\end{equation}
which corresponds to the excess energy with respect to the homogeneous
system, also called the distortion energy \cite{DeGennes}. For
$n=$constant, the integral in \myref{eq:surface-tension} is easily
expressed in terms of the mixing energy $W$:
\begin{equation}
\sigma= n \int_{\molf_A}^{\molf_B}d\molf\;
\sqrt{2 \kappa \widehat{W}(\molf)},
\label{eq:distortion-energy}
\end{equation}
with $\widehat{W}(\molf)=W(\molf)-W(\molf_A)>0$.
Without loss of generality, we can put $\kappa=k T_c
\xi^2$, so that the surface tension is $\sigma= k T_c n \xi \Omega$,
with $\Omega$ given by the integral
\begin{equation}
\Omega=  \int_{\molf_A}^{\molf_B}d\molf\;
\sqrt{2\widehat{W}(\molf)/k T_c}.
\label{eq:Omega}
\end{equation}
In the particular case $T=0$, the one--dimensional equilibrium molar fraction
is $\molf(x)=1/2+1/2 \sin(2 x/\xi)$ for $|x|<\pi \xi/2$, $0$ or $1$
otherwise. The interface width is $\pi \xi$ and the surface tension
integral~\myref{eq:Omega} gives $\Omega=\pi/4$. For $T>0$ the bulk
concentrations $\molf_A$ and $\molf_B$ of the flat interface profile are
approached asymptotically, and the exact definition of $\xi$ can be assigned
in an arbitrary way. In this work $\xi$ is defined as the interval over which
the flat interface concentration rises from $34.4\%$ to $65.6\%$ of the total
variation $\molf_B-\molf_A>0$. In other terms, given the profile $\molf(x)$,
$\xi=|x_2-x_1|$, where
$(\molf(x_1)-\molf_A)/(\molf_B-\molf_A)=1/2-1/2\sin(1/\pi)$,
$(\molf(x_2)-\molf_A)/(\molf_B-\molf_A)=1/2+1/2\sin(1/\pi)$.  In the
simulations reported here we chose $T/T_c=0.8$, such that the mixing
energy~\myref{eq:simple_mixture} is far from the singular limit $T=0$, and the
interface thickness is still quite small.

\subsection{Wetting}

In the framework of the Cahn--Hilliard theory of diffuse interfaces, the
interaction of the fluid components with a wall is introduced by adding to the
functional~\myref{eq:free_energy} a surface energy term of the
form~\cite{Cahn}
\begin{equation}
F_s[n_A,n_B]=\int_V \diffd\bfr\; \delta(S)\Phi(n_A,n_B),
\label{eq:Fs}
\end{equation}
where $\delta$ is Dirac delta function, $S(\bfr)=0$ defines the locus of the
points of the surface and $\Phi$ is the surface energy density.  From the
series expansion of $\Phi$~\cite{DeGennes}, we keep the first--order term
only, $\Phi=-\kappa (G_A n_A+G_B n_B)$, with the coefficients $G_A$, $G_B$
weighting the fluid--wall energy with respect to the same parameter $\kappa$
governing the fluid--fluid surface tension of
equation~\myref{eq:surface-tension}.  In the incompressible case, surface
energy variations depend only on the molar fraction $\molf$ and the difference
$G=G_B-G_A$. The addition of the surface free energy~\myref{eq:Fs} modifies
the Euler--Lagrange equation~\myref{eq:EulerLagrange} in the proximity of the
wall and imposes a discontinuity in the gradient of $\molf$:
\begin{equation}
-\kappa \nabla^2 \molf+W^\prime(\molf)-\kappa G \delta(S)=\mucoex;
\label{eq:EulerLagrange-wetting}
\end{equation}
the wall interaction is therefore included in the model as a boundary
condition by simply imposing the constraint
\begin{equation}
\mathbf{n}\cdot \nabla \varphi=G,
\label{eq:boundary1}
\end{equation}
where $\mathbf{n}$ is the normal to the wall surface, pointing inward to the
solid. The parameter $G$, which has dimensions of $1$/Length, determines the
wetting of the solid by the two phases; $G>0$ implies an attraction for
component B, $G<0$ a repulsion. As in the unbounded case of
equation~\myref{eq:EulerLagrange}, the properties of one--dimensional
solutions of equation~\myref{eq:EulerLagrange-wetting} can be deduced from the
first integral
\begin{equation}
\frac{\kappa}{2} \left[\frac{d \molf}{dx}\right]^2-\widehat{W}(\molf)=0. 
\label{eq:first-integral}
\end{equation}
In particular, the equilibrium values $\molf_s$ of the molar fraction
at the surface correspond to the solutions of
\begin{equation}
|G|= \sqrt{\frac{2}{\kappa}\widehat{W}(\molf)},
\label{eq:phis}
\end{equation}
and the solid--fluid surface energies $\sigma_{A}$ and $\sigma_{B}$ for the
two phases are obtained as the sum of $-\kappa n G\molf_s$ and the distortion
energy integrals of the from~\myref{eq:distortion-energy}
~\cite{DeGennes,Seppecher}. Together with the fluid--fluid surface tension
$\sigma$, the surface energies $\sigma_{A}$ and $\sigma_{B}$ determine the
static contact angle $\theta$ through Young's law,
$\cos \theta =(\sigma_A-\sigma_B)/\sigma$.

\subsection{Transport equations}

The transport current for the density $n_B$ ($n_A=n-n_B$, since we keep $n$
constant) is the sum of an advection and a dissipative term, there the latter
is proportional to the mobility $\Lambda$ and the chemical potential gradient
$\nabla\mu$. The fluid momentum equation, besides the non--linear advection
term, contains both reactive and dissipative forces, depending respectively on
the pressure tensor and the shear viscosity $\eta$:
\begin{eqnarray}
\frac{\partial n_B}{\partial t}&=& 
-\mathbf{\nabla} \cdot \left(n_B\mathbf{v}\right)+ \nabla
\cdot (\Lambda \nabla \mu) \\ \frac{\partial \rho \mathbf{v}}{\partial t}&=&
-\nabla \cdot \mathbf{P} -\nabla \cdot \left(\rho \mathbf{v} \mathbf{v}\right)+\nabla
\cdot\eta [\nabla\mathbf{v}+(\nabla\mathbf{v})^T ],
\end{eqnarray}
where $\rho$ is the mass density and $\mathbf{v}$ is the mass average
velocity.  The isotropic part of the pressure stress tensor
\myref{eq:pressure-tensor}, $p=n f^\prime(n)-f(n)$, is solved for by imposing
the constraint $\nabla \cdot \mathbf{v}=0$ rather than being derived by a
specific form of the bulk free energy of the system. We will also take equal
molar masses and a constant mobility for the two components, $\Lambda=D n/kT$,
with $D$ the diffusion coefficient.  The viscosity changes with the molar
fraction, $\eta=\eta^\ast h(\molf)$, and the following functional form is
assumed for $h$:
\begin{equation}
h(\molf)=\left\{
\begin{array}{cc}
\left(\frac{\lambda-1}{2\lambda} \right)
\tanh\left(\frac{\molf-1/2}{\chi} \right)
+ \frac{\lambda+1}{2 \lambda} 
& \textrm{if $\lambda\ge 1$},\\
\left(\frac{\lambda-1}{2} \right)\tanh\left(\frac{\molf-1/2}{\chi} \right)
+ \frac{\lambda+1}{2} & \textrm{if $\lambda < 1$ },
\end{array}
\right.
\label{eq:visratio}
\end{equation}
where $\lambda=\eta_B/\eta_A$ is the viscosity ratio and the parameter $\chi$
determines the sharpness of the transition of the viscosity as the phase
separating interface is crossed; when $\chi \ll \molf_B-\molf_A$, we have that
$\eta^\ast=\max\{\eta_A,\eta_B\}$ is the viscosity of the more viscous phase.

Given the unit time $\unit{T}$, length $\unit{L}$, mass $\unit{M}$, and
density $\unit{n}$, we have the following dimensionless groups: Peclet number
$\Peclet=\unit{L}^2/D \unit{T}$, Reynolds number $\Reynolds=\unit{\rho}
\unit{V} \unit{L}/\eta^\ast$ ($\unit{V}=\unit{L}/\unit{T}$), Weber number
$\Weber=\unit{\rho} \unit{V}^2 \unit{L}/\sigma$ and the Cahn number
$\Cahn=\xi/\unit{L}$. We will maintain the symbols $\mathbf{v}$ and $p$ of the
dimensionless fields ($p$ is rescaled with $\unit{M}/\unit{L}\unit{T}^2$), $W$
for the rescaled mixing energy $W \unit{T}^2/\unit{M}\unit{L}^2$, $t$ and
$\nabla$ for the dimensionless time and differential operator. The
dimensionless transport equations become
\begin{eqnarray}
\frac{\partial \molf}{\partial t}
&=& -\nabla \cdot (\molf
 \mathbf{v})+ \frac{1}{\Peclet} \nabla^2 \left[ - \Cahn^2 \nabla^2
\molf + W^\prime(\molf)
\right] \label{eq:ADV}
\\
\frac{\partial \mathbf{v}}{\partial t}&=&
-\nabla \cdot (\mathbf{v} \mathbf{v})
-\nabla p
-\frac{\Cahn}{\Weber\Omega} \nabla \cdot (\nabla \molf \nabla \molf)\\
&+&
\frac{1}{\Reynolds}
\nabla\cdot  h [\nabla\mathbf{v}+(\nabla\mathbf{v})^T ].
\label{eq:NS}
\end{eqnarray}
The phase--field Navier--Stokes equations converge to the classical sharp
interface behavior as the interface thickness, indicated by $\Cahn$, is
reduced to zero along with the diffusivity $1/\Peclet$~\cite{Jacqmin-NS}. The
mesh spacing used in the numerical solution imposes the main constraint on the
Cahn number, since a sufficient number of points is needed to describe the
interface profile in order to avoid spurious effects and to control the
convergence of the solver.  At finite $\Cahn$ straining flows can thin or
thicken the interface, and this must be resisted by diffusion, which therefore
requires $1/\Peclet$ to be high enough.  On the other hand, too large a
diffusion will overly damp the flow, which imposes a lower limit on the Peclet
number.  In our simulations, $\Cahn$ and $\Peclet$, rather than being referred
to the realistic values of an immiscible mixture such as water in oil, will
therefore be chosen to optimize the approximation of the sharp interface
physics, compatible with the limits imposed by the mesh size.  Despite the
fact that $\Cahn$ will be typically as large as $1/20$ and $\Peclet$ of the
order 10, which is determined by the fact that we have to deal with
three--dimensional simulations, we will see that the model is capable of
capturing the main features of droplet flow in simple microfluidic devices.


\section{The numerical algorithm}
\label{sec:method}

The differential equations~\myref{eq:ADV} and \myref{eq:NS} are discretized on
a uniform three--dimensional Cartesian grid with staggered velocities; the
molar fraction and pressure fields are collocated at the centers of the cubic
cells, while the velocity components are placed on the faces, and the
boundaries of the simulated domain always coincide with a face of a grid cell,
be it a wall, an inlet or a pressure outlet.  The time evolution is
implemented with an fully implicit Euler scheme, which is first order in
time. Given the array for the discretized fields at time step $n$,
$\phi^n_{i=0,\ldots,3}=\molf^n, v^n_1, v^n_2, v^n_3$ and $p^n$, their values
at time step $n+1$ satisfy a set of equations of the form
\begin{equation}
\phi^{n+1}_i=\phi^n-\Delta t 
N_i(\phi_1^{n+1},\phi_2^{n+1},\phi_3^{n+1},\phi_4^{n+1},p^{n+1}), 
\label{eq:scheme}
\end{equation}
with the constraint of incompressibility,
$\nabla\cdot\mathbf{v}^{n+1}=0$.  The explicit form of the nonlinear
functions $N_i$ is
\begin{eqnarray}
\lefteqn{N_0(\{\phi_i^{n+1}\},p^{n+1})=}\\
& &\disnab_\alpha (\molf^{n+1} v^{n+1}_\alpha)- \frac{1}{\Peclet} 
\left[- \Cahn^2 \disnab^4 \molf^{n+1} + \disnab^2 W^\prime(\molf^{n+1})
\right], \nonumber
\end{eqnarray}
\begin{eqnarray}
\lefteqn{N_\alpha(\{\phi_i^{n+1}\},p^{n+1})= }\\
& & \disnab_\beta (v^{n+1}_\alpha v^{n+1}_\beta)+\disnab_\alpha p^{n+1}
+\frac{\Cahn}{\Weber\Omega} 
\disnab_\beta (\disnab_\alpha \molf^{n+1} \disnab_\beta \molf^{n+1})\nonumber\\
& & -\frac{1}{\Reynolds}
\disnab_\beta [ h (\disnab_\alpha v^{n+1}_\beta+
\disnab_\beta v^{n+1}_\alpha) ]\hskip 1 cm \alpha=1,2,3, \nonumber
\end{eqnarray}
and the operators $\disnab_\alpha$ refer to the finite--difference
approximations of the partial derivative operators.  The iterative
scheme for the completion of the time step goes as follows:
\begin{enumerate}
\item Given the current guess of the new values for the fields,
$\phi_i^\ast$ and $p^\ast$, which at the beginning of the iteration
loop will be simply $\phi_i^{n}$ and $p^{n}$, the equations
\myref{eq:scheme} are solved in sequence to get the new guesses
$\phi_i^{\ast\ast}$
\begin{eqnarray}
\molf^{\ast\ast}&=&\molf^n-
\Delta t N_1(\molf^{\ast\ast},v_1^\ast,v_2^\ast,v_3^\ast,p^\ast),\\
v_1^{\ast\ast}&=&v_1^n-
\Delta t N_2(\molf^{\ast\ast},v_1^{\ast\ast},v_2^\ast,v_3^\ast,p^\ast),\\
v_2^{\ast\ast}&=&v_2^n -
\Delta \tau N_3(\molf^{\ast\ast},v_1^{\ast\ast},v_2^{\ast\ast},
v_3^\ast,p^\ast),\\
v_3^{\ast\ast}&=&v_3^n -\Delta \tau N_4(\molf^{\ast\ast},v_1^{\ast\ast},
v_2^{\ast\ast},v_3^{\ast\ast},p^\ast).
\end{eqnarray}
Each inner iteration is solved recursively by substituting
$\phi^{\ast\ast}\simeq \phi_{k+1}=\phi_k+\delta_k$ with
$\phi_0=\phi^n$ and linearizing in $\delta_k$.
\item the continuity equation, is enforced with a
prediction--correction method (see, for example, ref.~\cite{Wesseling}), leading to new values 
$v_i^{\ast\ast\ast}$ and $p^{\ast\ast}$,
\item the outer iteration is completed by setting
$\molf^{\ast\ast}\rightarrow\molf^\ast$,
$v^{\ast\ast\ast}_i\rightarrow v^\ast_i$, $p^{\ast\ast}\rightarrow
p^\ast$, and computing the of residuals
\begin{equation}
R_i(\phi_i^\ast)=\phi_i^n-\phi_i^\ast-N_i(\{ \phi_i^\ast\},p^\ast),
\hskip 0.2 cm i=0,1,2,3.
\end{equation}
together with the divergence field $D=\disnab \cdot u^{\ast}$; the
outer iteration loop is completed if $\norm{R_0(\molf^\ast)}< \epsilon
\norm{\molf^\ast}$ and $\norm{R_i(\molf^\ast)}< \epsilon
\norm{v^\ast}$, $i=1,2,3$, and $\norm{D}< \epsilon$, all satisfied.
In all simulations presented in this paper we set
$\epsilon=10^{-5}$. When the desired accuracy is reached, the new
fields are obtained by setting $\phi_i^{n+1}=\phi_i^{\ast}$,
$p^{n+1}=p^{\ast}$.
\end{enumerate}
Besides the periodic boundaries used for the case of unbounded flow, we can
handle walls, inlet and outlets. For all three kind of boundaries the normal
gradient of the chemical potential is set to zero ($\mathbf{n}\cdot \nabla
\mu=0$). The boundary conditions (BC) at the walls include the
constraint~\myref{eq:boundary1} for the molar fraction, the no--flow
conditions $\mathbf{n}\cdot \nabla p=0$ and $\mathbf{v}_\perp=0$, and the
no--slip condition $\mathbf{v}_\parallel=\mathbf{v}_\mathrm{wall}$, where
$\mathbf{v}_\mathrm{wall}$ is the tangential velocity of the wall.  Inlets
have Dirichlet BC for the molar fraction and velocity and Neumann BC for the
pressure ($\molf=\mathrm{const.}$, $\mathbf{v}_\perp=\mathrm{const.}$,
$\mathbf{v}_\parallel=0$, $\mathbf{n}\cdot \nabla p=0$), while outlets have
the complementary setup ($\mathbf{n}\cdot \nabla\molf=0$, $\mathbf{n}\cdot
\nabla\mathbf{v}=0$, $\mathbf{v}_\parallel=0$, $p=\mathrm{const.}$).

\section{Free flow dynamics}
\label{sec:free_flow}

\subsection{Spurious currents}

\begin{figure*}[tbp]
\begin{center}
\includegraphics[width=0.28\textwidth, angle = -90]{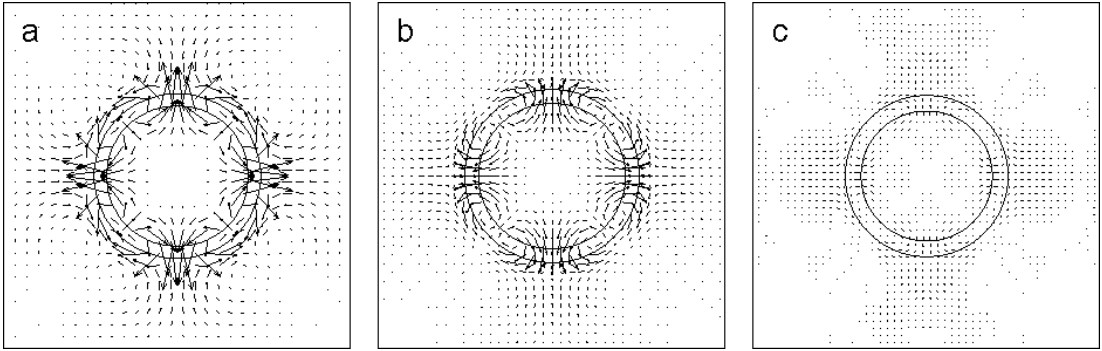}
\end{center}
\caption{Visualization of spurious currents for a static droplet relaxed on
  three different cubic meshes: a) $\Delta x=\Cahn$, b) $\Delta x=2/3 \Cahn$,
  c) $\Delta x=1/2 \Cahn$, $\Cahn=0.1$. The pressure, concentration and
  velocity fields have reached the stationary state, and parasite currents
  persist in the proximity of the interface region indicated by the concentric
  circles. The spurious currents reflect the four--fold symmetry of the
  lattice, and depend on the size of the mesh. Measured in units of the
  reference velocity $\Reynolds/\Weber$, the maximum spurious velocity
  decreases from $4\cdot 10^{-3}$ (a) to $2 \cdot10^{-3}$ (b), and $3\cdot
  10^{-4}$ (c).}
\label{fig:spurious}
\end{figure*}

The appearance of spurious currents in the proximity of the phase boundaries
is a common feature shared by all numerical methods which include capillary
stresses in the Navier--Stokes equation as a volume force to be computed on a
grid~\cite{ScardovelliAnnRevFluidMech1999}. A simple numerical test to
illustrate their characteristics is to consider a droplet at rest; when the
pressure and molar fraction fields are equilibrated, the velocity field is not
zero, and parasite currents appear in the interface region, with a pattern
which reproduces the symmetry of the underlying lattice
(figure~\ref{fig:spurious}). The origin of the spurious currents can be
explained by analyzing how the discretized form of the differential operators
modifies the continuum transport equations (see Appendix), and their magnitude
grows with the coarseness of grid and the ratio between surface tension and
viscosity. In our simulations the size of spurious currents was controlled by
changing the lattice spacing, keeping the modulus below $5\%$ of the
characteristic velocity of the flow.

\subsection{Droplet deformation and breakup}

\begin{figure}[htb]
\begin{center}
\includegraphics[width=0.32\textwidth, angle = -90]{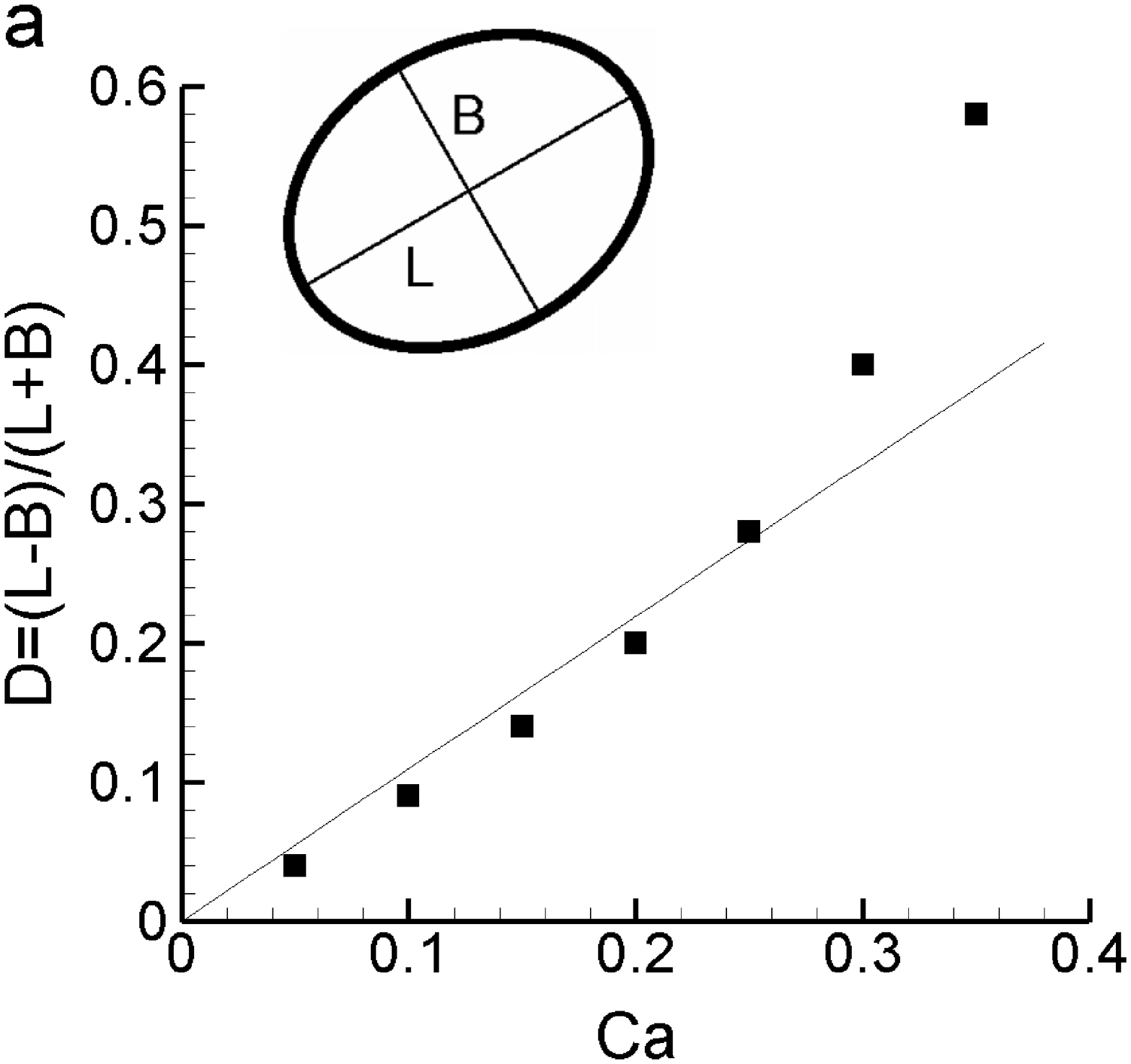}
\includegraphics[width=0.32\textwidth, angle = -90]{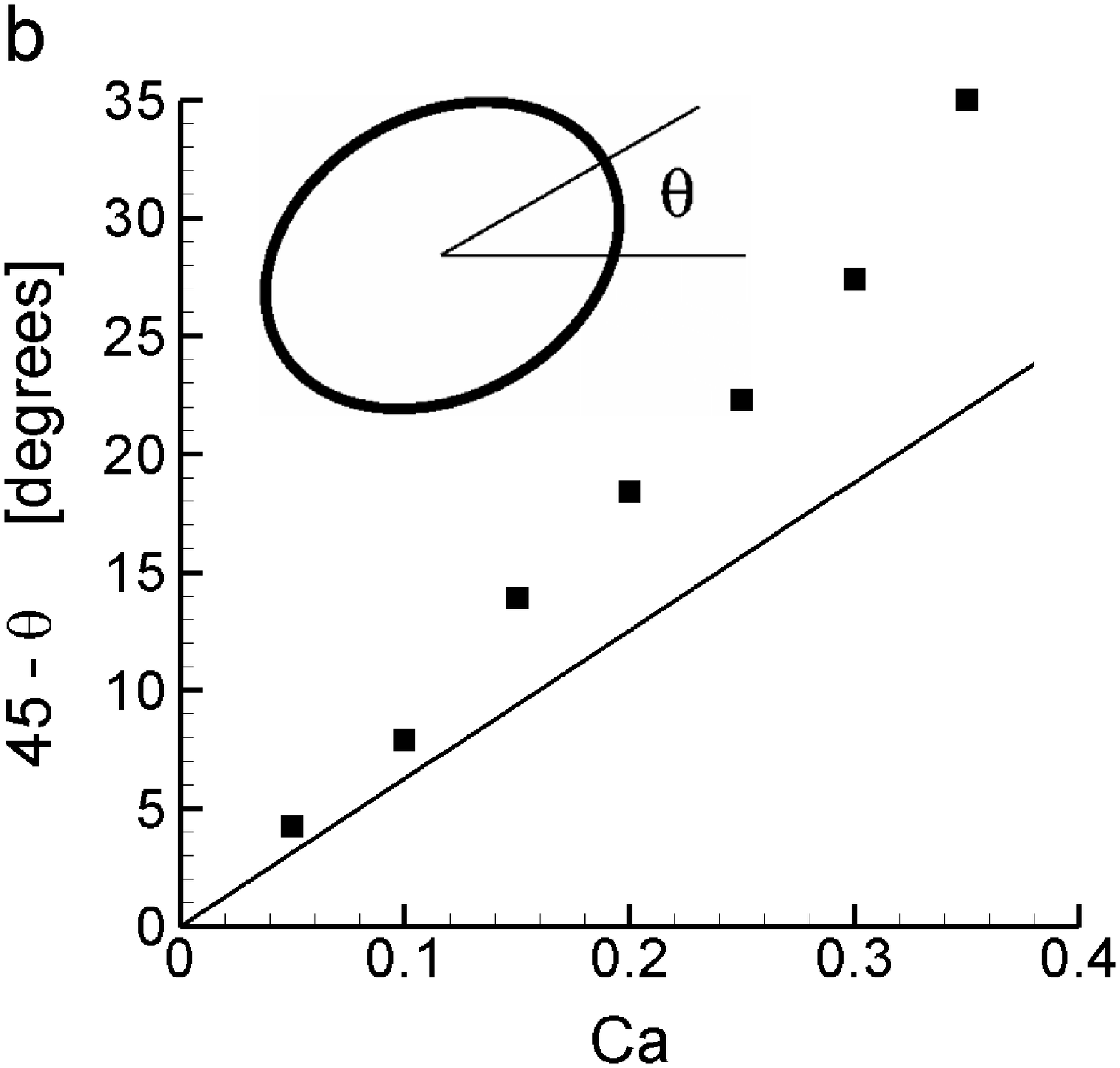}
\end{center}
\caption{Variation of deformation (a) and tilt angle (b) as a function of
capillary number for a droplet in simple shear. The straight lines show the
$\Capillary \ll 1$ limits derived by Taylor~\cite{Taylor} and Chaffey and
Brenner~\cite{Chaffey} ($\lambda=1$). The simulations were performed on a
$80\times60\times60$ box with the top and bottom faces moving with opposite
velocities parallel to the longer side. $\Cahn=\Delta x = 0.05$. }
\label{fig:deformation_angle}
\end{figure}

Droplet deformation and breakup in unbounded simple shear flow depends on the
ratio between droplet and matrix viscosities, $\lambda=\eta_d/\eta_c$, and the
capillary number $\Capillary = \eta_c \shear R/\sigma$ where $\shear$ is the
shear rate and $R$ is the radius. For low $\Capillary$, surface stresses are
balanced by viscous ones and the droplet is stretched to a stationary
ellipsoidal shape, with its principal axis following the direction of maximum
elongation; the deformation is expressed in terms of Taylor's parameter
$D=(L-B)/(L+B)$~\cite{Taylor}, $L$ being the long breadth of the drop, and $B$
the small breadth (see inset to figure~\ref{fig:deformation_angle}a). The
angle $\theta$ between the axis of maximum elongation and the shear direction
gives the orientation of the droplet. Simulation results are consistent with
small deformation theories showing a linear relation between $D$ and $\theta$
and $\Capillary$ for $\Capillary\ll1$ (figure~\ref{fig:deformation_angle}b).
As the capillary number is increased above a critical value
$\CapillaryCritical$, no steady shape exists and the droplet breaks.  In our
simulations, we find
$0.4<\CapillaryCritical<0.44$, in agreement with the value
$\CapillaryCritical\approx 0.41$ obtained with more accurate numerical
methods~\cite{Rallison1981,Li}.
%

%

\section{Droplet flow in square microchannels}
\label{sec:channels}

\begin{figure}[htb]
\begin{center}
\includegraphics[width=0.4 \textwidth, angle = -90]{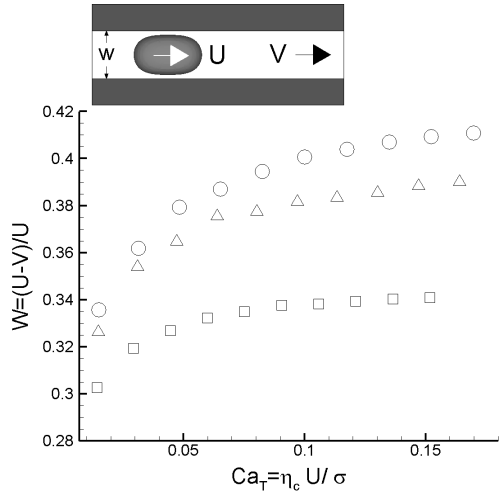}
\end{center}
\caption{Droplet flowing in a capillary with square cross section of width
  $w$; the (normalized) droplet excess speed is plotted against the capillary
  number for different viscosities: $\lambda=1/32$ ($\bigcirc$), $\lambda=1/8$
  ($\bigtriangleup$), and $\lambda=1/2$ ($\Box$). $U$ is the droplet speed
  while the $V$ is average flow speed of the continuous phase (see inset). The
  undeformed radius of the droplet is $0.55 w$. The simulation mesh is made by
  $125\times25\times25$ points, $\Cahn=\Delta x = 0.04$. }
\label{fig:droplet_pipe}
\end{figure}

As a first example of two--phase flow in a constrained geometry we discuss the
the pressure driven motion of droplets inside a square channel. By choosing
the appropriate value of the parameter $G$
(equation~\myref{eq:EulerLagrange-wetting}), the contact angle to the wall is
set to zero, so that the continuous phase completely wets the boundaries. Exact
results for droplet motion in capillaries are known only in the case of
circular tubes~\cite{Hodges2004}. In this case, theory shows that the speed of
the drop $U$ exceeds the average fluid speed by and amount $UW$, where
$W\propto \CapillaryTube^{2/3}$, and $\CapillaryTube=\eta_c U/\sigma$ is the
capillary number. In the case of square channels, to our knowledge,
theoretical and experimental studies have been published only on the flow of
bubbles, that is in the limit of small droplet viscosity $\eta_d\to
0$~\cite{Kolb1993,WongBubblesI1995,WongBubblesII1995,Thulasidas1995}.  For
viscous drops in square channels we verified that the value of $\lambda$
strongly influences the droplet speed; at a given $\CapillaryTube$ the
fractional velocity correction W increases as the viscosity ratio is
decreased. We tested in particular the motion of droplets whose diameter is
comparable with the channel width (figure~\ref{fig:droplet_pipe}), and
verified that the values of $W$ in the range $0.01<\CapillaryTube<0.2$ are
consistent with the experimental results for bubbles by Thulasidas et
al.~\cite{Thulasidas1995}, i.e. $0.3<W<0.45$.

\section{Droplet breakup in a T--junction}
\label{sec:Tjunction}

\begin{figure*}[htb]
\begin{center}
\includegraphics[width=0.62\textwidth, angle = -90]{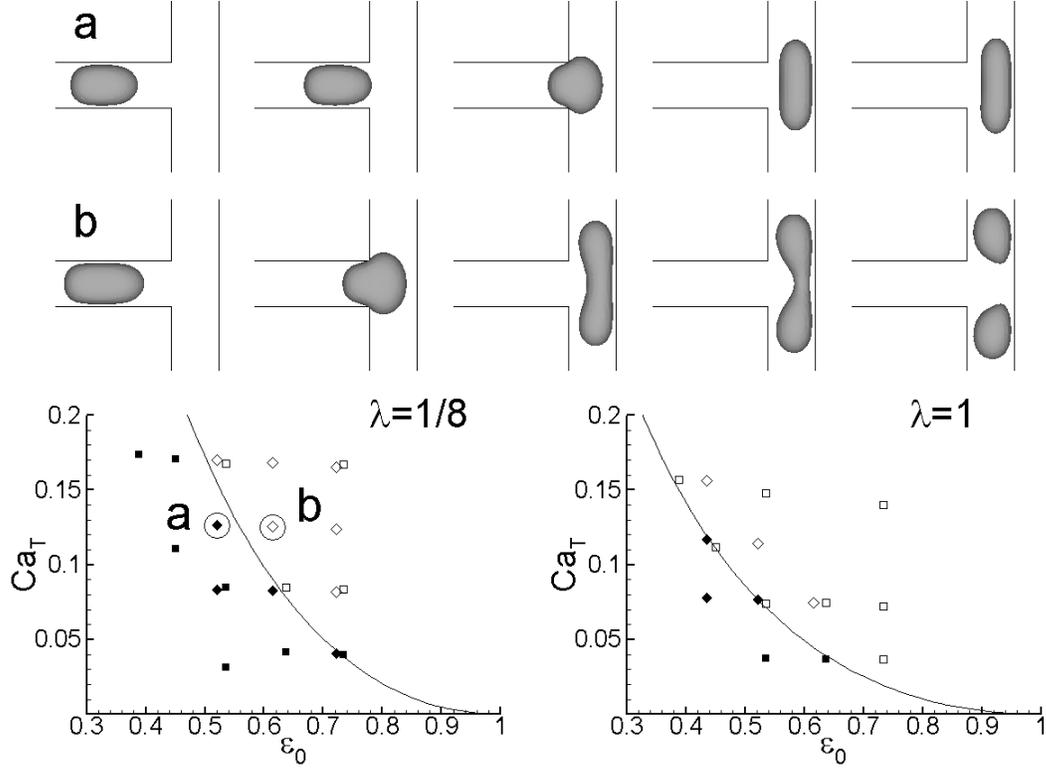}
\end{center}
\caption{Bottom: $(\varepsilon_0,\CapillaryTube)$ diagram summarizing the
  conditions for breaking drops at a symmetric T--junction for $\lambda=1/8$
  (left) and $\lambda=1$ (right). Empty symbols correspond to non--breaking
  droplets, filled symbols to breaking droplets; diamonds correspond to
  simulations preformed with $\Cahn=1/15$, while squares refer to simulations
  with $\Cahn=1/20$. The solid lines show the separating boundary from
  equation~\myref{eq:critical_Ca}, with the viscosity dependent coefficients
  $\alpha(1/8)=1$ and $\alpha(1)=0.5$. Top: the frames in a) and b) show the
  motion of droplets with different elongation and same capillary number,
  corresponding to the points labeled a and b in the phase diagram for
  $\lambda=1/8$.}
\label{fig:Tdiagram}
\end{figure*}

These simulations were motivated by the experiments by Link et al.~\cite{Link}
on droplet breakup in a T--shaped junction
configuration~\cite{Thorsen}. Isolated water droplets dispersed in an oil
matrix are transported by a pressure--driven flow along a hydrophobic
microchannel until they reach the junction, where they are stretched by the
local elongational flow. Depending on the droplet size and relative strengths
of viscous to capillary stresses, they split or simply flow down one of the
two branches (see insets a) and b) in figure~\ref{fig:Tdiagram}).  The
dimensionless groups are the capillary number $\CapillaryTube$, the viscosity
ratio, $\lambda=\eta_d/\eta_c$ and the initial droplet extension
$\varepsilon_0=l_0/\pi w_0$, where $l_0$ and $w_0$ are the length and the
diameter respectively of the droplet upstream of the T--junction.

Figure~\ref{fig:Tdiagram} summarizes the results for two values of $\lambda$
by reporting for each simulation a point with coordinates
$(\varepsilon_0,\CapillaryTube)$; empty and filled symbols mark breaking and
non--breaking droplets respectively. These two regions are separated by a
curve curve whose shape can be derived with a general argument based on the
Rayleigh--Plateau limit~\cite{Link}, and is given by the function
\begin{equation}
\CapillaryCritical=\alpha
\varepsilon_0 (1/\varepsilon_0^{2/3}-1)^2,
\label{eq:critical_Ca}
\end{equation}
 where $\alpha(\lambda)$ is a constant which depends on the viscosity
contrast. We verified that this function describes very accurately the
critical line for a range of $\lambda$.  In particular we get
$\alpha(1/8)=1.02\pm0.03$, $\alpha(1/2)=0.60\pm0.04$, $\alpha(1)=0.50\pm0.01$,
$\alpha(3/2)=0.35\pm0.04$, $\alpha(2)=0.33\pm0.04$.  These data clearly
indicate that droplet breakup in the T--junction occurs more easily as the
viscosity ration is increased, even when $\lambda>1$. The variation is quite
rapid as the region around $\lambda=1$ is crossed, reaching a uniform value
for large $\lambda$.  Such behavior is in agreement with the viscosity
dependence of the critical capillary number for a free droplet in a purely
elongational flow~\cite{BentleyDropletBreakup1986}. Our result for
$\alpha(1/8)$ closely agrees with the experimental value ($\alpha=1$) obtained
with water droplets dispersed in hexadecane oil~\cite{Link} (viscosity ratio
$\lambda=1/8$), and is a demonstration of the accuracy of the numerical method
presented here. As a final remark on the computational side, we would like to
point out that the position of the critical line remains quite consistent for
different values of the Cahn number; figure~\ref{fig:Tdiagram} in fact reports
data obtained for $\Cahn=1/15$ and $\Cahn=1/20$, showing how such relatively
large values are sufficient to reproduce the critical boundary described by
the function~\myref{eq:critical_Ca}.
%

\section{Conclusions}

We have discussed the application of a phase--field method to the modeling of
 two--phase flow in microfluidic devices in three dimensions; in particular,
 this approach is validated by investigating the dynamics of droplet breakup
 in a symmetric T--shaped junction. The phase diagram obtained from the
 numerical simulations agrees very closely with the experimental results which
 were performed with water--in--oil dispersions, confirming the reliability of
 the numerical approach. This conclusion was not guaranteed, given that the
 physics of the binary mixture approaches that of two immiscible fluids only
 in the limit of an interface thickness which is vanishingly small compared to
 the droplet size. Also, we were able to show that the resistance of the
 droplet to breakup depends on the viscosity ratio of the two fluids,
 following the characteristic trend of isolated free droplets in a purely
 elongational flow.

 The main advantage of the method stems exactly from the fact that surface
 stresses are distributed over the finite interface thickness, allowing
 coarser meshes and therefore shorter simulation times. Furthermore, compared
 to other diffuse--interface models, the transport equations are derived from
 a generalized free--energy functional, which determines also the equilibrium
 properties of the system, and is the starting point for extending the model
 to include more components and phases.  The efficiency of the method, which we
 have shown does not come at the expense of accuracy or reliability, offers
 the possibility of handling two--phase flow in complex microfluidic devices;
 at the fundamental level, this should have an impact in the understanding of
 phenomena related to pattern formation in such devices, and we may expect
 that implementations of the method will prove to be useful for design
 purposes.

\begin{acknowledgements}
The author  would like to thank W.~G.~M. Agterof for useful discussions. The
development of the code was supported by a Marie Curie Industrial Host
Fellowship (HPMI-CT-1999-00017).  Support form Unilever Corporate Research and
the European Union (MRTN-CT-2004-005728) is also acknowledged. Special thanks
to H.~A. Stone and F. Jousse for their comments on the manuscript.
\end{acknowledgements}

\appendix*
\section{Spurious currents}

The origin of spurious currents can be explained by analyzing how the
discretized form of differential operators modifies the continuum transport
equations~\myref{eq:ADV} and~\myref{eq:NS}. Given set of stencil vectors
$\{\stene_i\}$ and the lattice spacing $\Delta x$, the action of
$\disnab_\alpha$ and $\disnab^2$ on a field $f(\bfr)$ can be written in the
general form
\begin{eqnarray}
\disnab_\alpha f(\bfr)&=&\frac{1}{2\Delta x}\sum_i t_i f(\bfr+\Delta x \stene_i) \stene_{i\alpha}
\label{eq:discrete_grad}
\\ \disnab^2 f(\bfr)&=&\frac{1}{(\Delta x)^2}\sum_i t_i f(\bfr+\Delta x \stene_i),
\label{eq:discrete_lapl}
\end{eqnarray}
where the coefficients $\{t_i\}$ are chosen to ensure the equality between the
discretized and continuum operators to the leading order. Their values are
obtained by expanding in Taylor series the term $f(\bfr+\Delta x \stene_i) $
in equations~\myref{eq:discrete_grad} and~\myref{eq:discrete_lapl}; for a
stencil having inversion symmetry, the expansion contains only the lattice
tensors with an even number of indices
\begin{eqnarray}
T^{(2)}_{\alpha\beta}&=&\sum_i t_i \stene_{i\alpha}\stene_{i\beta}, \\
T^{(4)}_{\alpha\beta\gamma\delta}&=&\sum_i t_i
\stene_{i\alpha}\stene_{i\beta}\stene_{i\gamma}\stene_{i\delta}, 
\\
 \ldots \nonumber
\end{eqnarray}
and the effect of the anisotropy of the lattice on the actual expression of
the differential operators depends on the invariance properties of the tensors
$T^{(n)}$.  In the case of the simple cubic stencil with the $6+1$ base
vectors $\{\stene_i\}=(\pm 1,0,0), (0,\pm 1,0), (0,\pm 1,0)$ and
$\stene_0=(0,0,0)$, the values $t_0=0$ for the gradient, $t_0=-6$ for the
Laplacian, and $t_{i=1,\ldots,6}=1$, imply that $T^{(2)}_{\alpha\beta}=2
\delta_{\alpha\beta}$ is isotropic, such that $\disnab_\alpha
f(\bfr)=\nabla_\alpha f(\bfr)+O(\Delta x^2)$, and $\disnab^2 f(\bfr)=\nabla^2
f(\bfr)+O(\Delta x^2)$. The coefficients of the fourth order tensor $T^{(4)}$,
on the other hand, depend on the orientation of the stencil vectors with
respect to the coordinate axis, and are responsible for the anisotropy effects
in the discretized transport equations.

We now give a simple qualitative argument to relate the magnitude of the
 parasite currents to the dimensionless groups $\Weber$, $\Reynolds$ and
 $\Cahn$ and the lattice spacing $\Delta x$.  Let us consider the example of
 the equilibrated droplet in absence of any external flow. The transport
 equations~\myref{eq:ADV} and~\myref{eq:NS} are simplified to $\mu=$ const.,
 $\nabla_\beta P_{\alpha\beta}=0$ and $\mathbf{v}=0$; in particular, the
 components $F_\alpha$ of the volume force are
\begin{equation}
F_\alpha=\nabla_\alpha p + \frac{\Cahn}{2 \Weber\Omega} \nabla_\alpha
|\nabla\molf|^2 + 
\frac{\Cahn}{\Weber\Omega} \nabla^2\molf \nabla_\alpha\molf=0,
\end{equation}
or
\begin{equation}
F_\alpha=\nabla_\alpha \ptilde + \frac{\Cahn}{\Weber\Omega} \nabla^2\molf
\nabla_\alpha\molf=0,
\label{eq:volume_force}
\end{equation}
where we have redefined the isotropic pressure in the interface region,
$\ptilde=p+ (|\nabla\molf|^2/2) (\Cahn/\Weber\Omega)$. The discretized form of
the volumetric force $F_\alpha$ is obtained with the replacement
$\nabla_\alpha\rightarrow\disnab_\alpha$, $\nabla^2\rightarrow\disnab^2$ in
equation~\myref{eq:volume_force}. When the fields $\ptilde^{(0)}$ and
$\molf^{(0)}$ solving equation~\myref{eq:volume_force} for the equilibrated
droplet are collocated on a grid, there is a residual stress of order
$O(\Delta x^2)$ which depends explicitly on $T^{(4)}$:
\begin{eqnarray}
\lefteqn{\widehat{F}_\alpha[\ptilde^{(0)},\molf^{(0)}]=}\nonumber\\
& & \Delta x^2\left [
\frac{ T^{(4)}_{\alpha\beta\gamma\delta}}{2\cdot 3! }
 \partial_{\beta\gamma\delta}\ptilde^{(0)}
+\frac{\Cahn}{\Weber\Omega} (\partial_\alpha \molf^{(0)})
\frac{T^{(4)}_{\beta\gamma\delta\epsilon}}{4! } 
\partial_{\beta\gamma\delta\epsilon}\molf^{(0)}
\right]\nonumber\\
& & +O(\Delta x^4)\neq 0
\end{eqnarray}
and is balanced (neglecting the non--linear advection term in~\myref{eq:NS})
by the shear stress produced by a non--zero velocity field $\mathbf{v}^{(1)}$,
such that $\widehat{F}_\alpha= \nabla^2 v^{(1)}_\alpha/\Reynolds$.  For the
case of equal viscosities, given that $\ptilde^{(0)}$ in the interface region
is proportional to $1/(\Weber \Cahn)$ and that each partial derivative
introduces a factor $1/\Cahn$, the module of the spurious currents is
estimated by the simple relation
\begin{equation}
 v^{(1)}\propto \left(\frac{\Delta x}{\Cahn} \right)^2
\frac{\Reynolds}{\Weber},
\label{eq:spurious}
\end{equation}
and the proportionality factor depends on the non--invariant tensor $T^{(4)}$
which is responsible for the angular dependence of the parasite currents along
the surface of the static droplet. Equation~\myref{eq:spurious} includes the
main factors governing the the magnitude of spurious currents, which are small
for low surface tension or high viscosity, and can be controlled by including
more grid points in the interface region, therefore decreasing the ratio
$\Delta x/\Cahn$. This option is the most straightforward to use, at the cost
of larger grids and therefore longer computational times. A more effective
possibility consists in improving the symmetry properties of the
discretization stencil $\{\stene_i\}$ by adding more vectors, such that the
$T^{(4)}$ becomes isotropic and corrections are reduced to order $(\Delta
x/\Cahn)^4$~\cite{Jacqmin-NS}.

\end{document}